\documentclass{aastex}
\usepackage{graphicx}
\usepackage[space]{grffile}
\usepackage{latexsym}
\usepackage{textcomp}
\usepackage{longtable}
\usepackage{multirow,booktabs}
\usepackage{amsfonts,amsmath,amssymb}
\usepackage{natbib}
\usepackage{url}
\newif\iflatexml\latexmlfalse
\usepackage[utf8]{inputenc}
\usepackage[ngerman,english]{babel}

\shorttitle{TRIPPy: Trailed Image Photometry in Python}
\shortauthors{Wesley Fraser}

\begin{document}


\title{TRIPPy: Trailed Image Photometry in Python}


\author{Wesley Fraser}
\affil{Queen's University Belfast, Astrophysics Research Center, David Bates Building, Belfast, United Kingdom, BT7 1NN}
 
\author{Mike Alexandersen}
\affil{Institute for Astronomy and Astrophysics, Academia Sinica}
 
\author{Megan E. Schwamb}
\affil{Institute for Astronomy and Astrophysics, Academia Sinica}
 
\author{Michaël Marsset}
\affil{European Southern Observatory, Aix Marseille University, CNRS, LAM}
 
\author{Rosemary E. Pike}
\affil{University of Victoria}
 
\author{JJ Kavelaars}
\affil{NRC Herzberg, National Research Council, Canada}
 
\author{Michele T. Bannister}
\affil{University of Victoria}
 
\author{Susan Benecchi}
\affil{Planetary Science Institute}
 
\author{Audrey Delsanti}
\affil{Aix Marseille University, CNRS, LAM}




\begin{abstract}
Photometry of moving sources typically suffers from reduced signal-to-noise (SNR) or flux measurements biased to incorrect low values through the use of circular apertures. To address this issue we present the software package, TRIPPy: TRailed Image Photometry in Python. TRIPPy introduces the pill aperture, which is the natural extension of the circular aperture appropriate for linearly trailed sources. The pill shape is a rectangle with two semicircular end-caps, and is described by three parameters, the trail length and angle, and the radius. The TRIPPy software package also includes a new technique to generate accurate model point-spread functions (PSF) and trailed point-spread functions (TSF) from stationary background sources in sidereally tracked images. The TSF is merely the convolution of the model PSF, which consists of a moffat profile, and super sampled lookup table. From the TSF, accurate pill aperture corrections can be estimated as a function of pill radius with a accuracy of 10 millimags for highly trailed sources. Analogous to the use of small circular apertures and associated aperture corrections, small radius pill apertures can be used to preserve signal-to-noise of low flux sources, with appropriate aperture correction applied to provide an accurate, unbiased flux measurement at all SNR.
\end{abstract}%

\bibliographystyle{apj}

\section{Introduction \label{sec:Intro}}

The brightness of the Solar System's moving bodies is one of the few directly measurable properties of those objects, and can provide insight into a body's size, shape, and surface properties. The primary challenge to accurate photometry of moving targets is a result of the apparent motion of the targets themselves. Typically, sources are observed with the telescope tracked either sidereally, or at the apparent motion of the target of interest. Either case results in the obfuscation of the target's trailed point-spread function (hereafter TSF), and consequently, a reduction in the overall photometric signal-to-noise (SNR) compared to equivalently bright stationary sources.  

Historically, photometry of faint trailed sources is done in the same way as for stationary sources, through the use of a small circular aperture, and associated aperture correction. This procedure suffers from two separate issues. The first issue is the obvious - but usually neglected - loss of flux due to trailing in sidereal images which is not accounted for by the aperture correction measured from stationary sources. This results in an under-reporting of the source's true flux. This issue can be avoided with the use of an aperture large enough to encompass the entire source image, but results in the second issue with the use of circular apertures - a large penalty in SNR as a result of inclusion of a larger background region. The scale of both effects is presented in Figure~\ref{fig:trailing}. 

Recently, \citet{Veres2012} presented a technique for trail fitting of moving sources. Aimed primarily at accurate astrometry, their method achieved this by fitting an axisymmetric gaussian profile convolved with a line to the trailed source. The apparent rate and angle of motion of a source was solved for, thereby providing extremely accurate astrometry. As we will discuss however, without the use of an associated lookup table, photometry determined from this PSF fitting method can be inaccurate.

Here we present a new pill-shaped aperture and software package, TRIPPy, designed specifically for precision photometry of both moving and stationary sources in sidereally tracked images. Through the use of an elongated pill-shaped aperture (a rectangle with two semicircular end-caps described by three parameters, the trail length and angle, and the radius), and an estimate of the TSF through convolution of the stationary source PSF, associated aperture corrections can be accurately estimated from the TSF, largely mitigating the issues associated with circular aperture photometry. The routines for both TSF generation, as well as pill and circular aperture photometry are available as a python package \citep{fraserw_48694}.\footnote{\url{https://github.com/fraserw/TRIPPy}} TRIPPy is a complete photometry package, which includes facilities for generation of the PSF and TSF, and mechanisms for photometry using PSF and TSF source fitting. As well, TRIPPy includes facilities for pill and circular aperture photometry, and background estimation through various techniques, including a new modal estimation technique we present here for measuring the background in crowded fields.  We present the aperture shape in Section 2, the recipe for PSF and TSF generation and resultant aperture correction quality in Section 3, and finish with concluding remarks in Section 4.

\section{Aperture Comparison\label{sec:aperture}}
\begin{figure}[h!]
\begin{center}
\includegraphics[width=0.7\columnwidth]{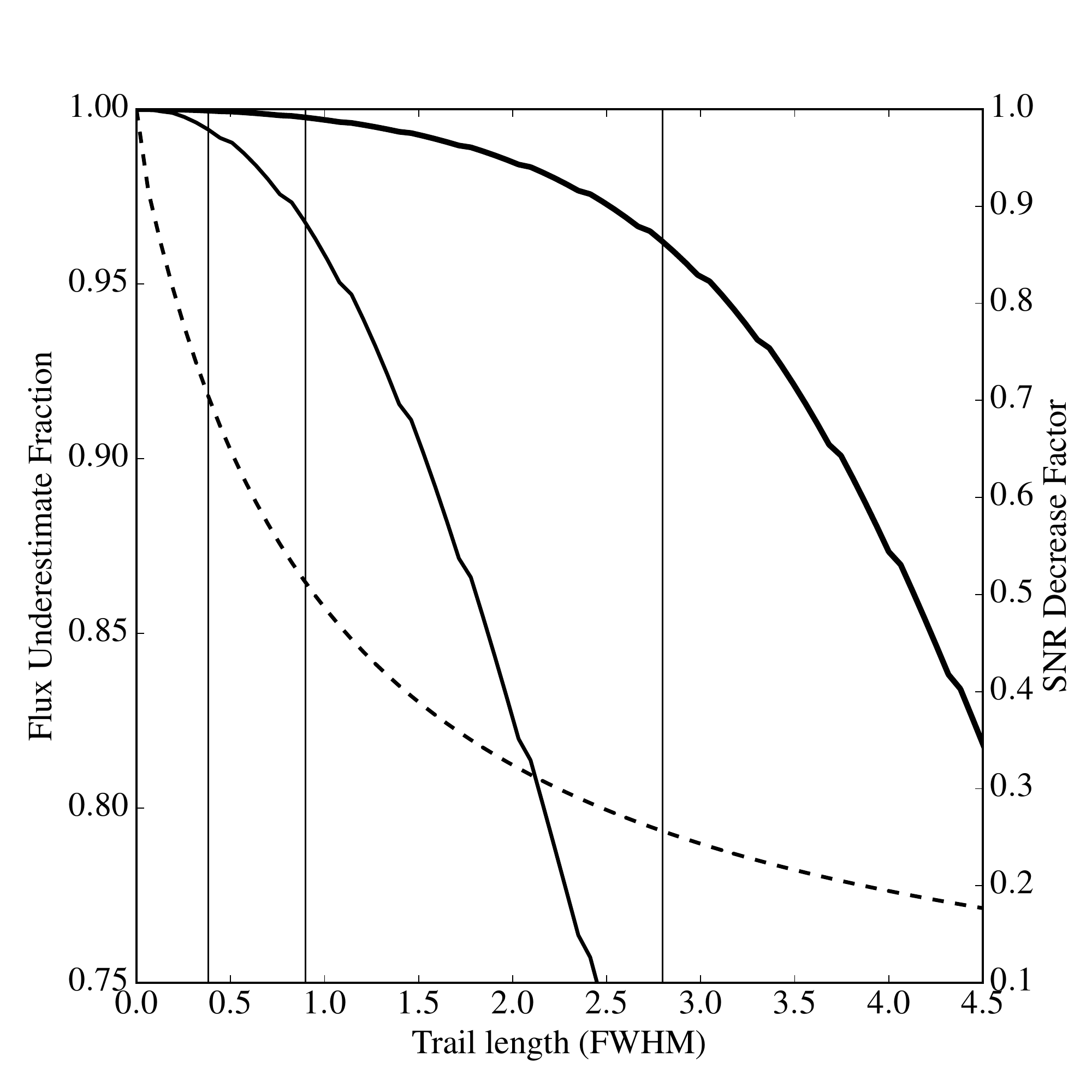}
\caption{\label{fig:trailing} Flux loss of a trailed source measured with circular apertures of radii 1 and 2 FWHM (solid thin and thick curves) vs. trailing length in units of the image FWHM. The dashed line (right ordinate) presents the SNR ratio between the trailed source, and a circular equivalent, both measured with circular apertures large enough to encompass $>99\%$ of the source flux. The three vertical lines represent the trailing of objects at 5, 16 and 40 AU, observed at opposition in 0.7" seeing with a 300~s exposure. The unevenness of the flux loss curves are due to pixel noise of the artificial images from which those curves were generated.%
}
\end{center}
\end{figure}

To understand the issues associated with circular aperture photometry of trailed sources, in Figure~\ref{fig:trailing} we present the fractional flux excluded from circular apertures of 1 and 2 full-widths at half maxima (FWHM) versus source trail length. In addition, we present the effective background-limited decrease in photometric precision of a trailed source compared to the stationary equivalent when an aperture large enough to include $>99\%$ of the source flux is utilized. The curves in Figure~\ref{fig:trailing} were measured from ideal trailed sources of various lengths, generated from a moffat profile with FWHM=5.5~pixels.

As can be seen, as long as the aperture radius is at least as large as the trailing length in the image, the flux excluded by the use of a circular aperture is no more than $\sim5\%$, a loss that can be effectively ignored for low SNR ($\lesssim15$) measurements. Higher precision photometry however, is possible with knowledge of the TSF, and a reasonable choice in aperture shape.

\begin{figure}[h!]
\begin{center}
\includegraphics[width=0.7\columnwidth]{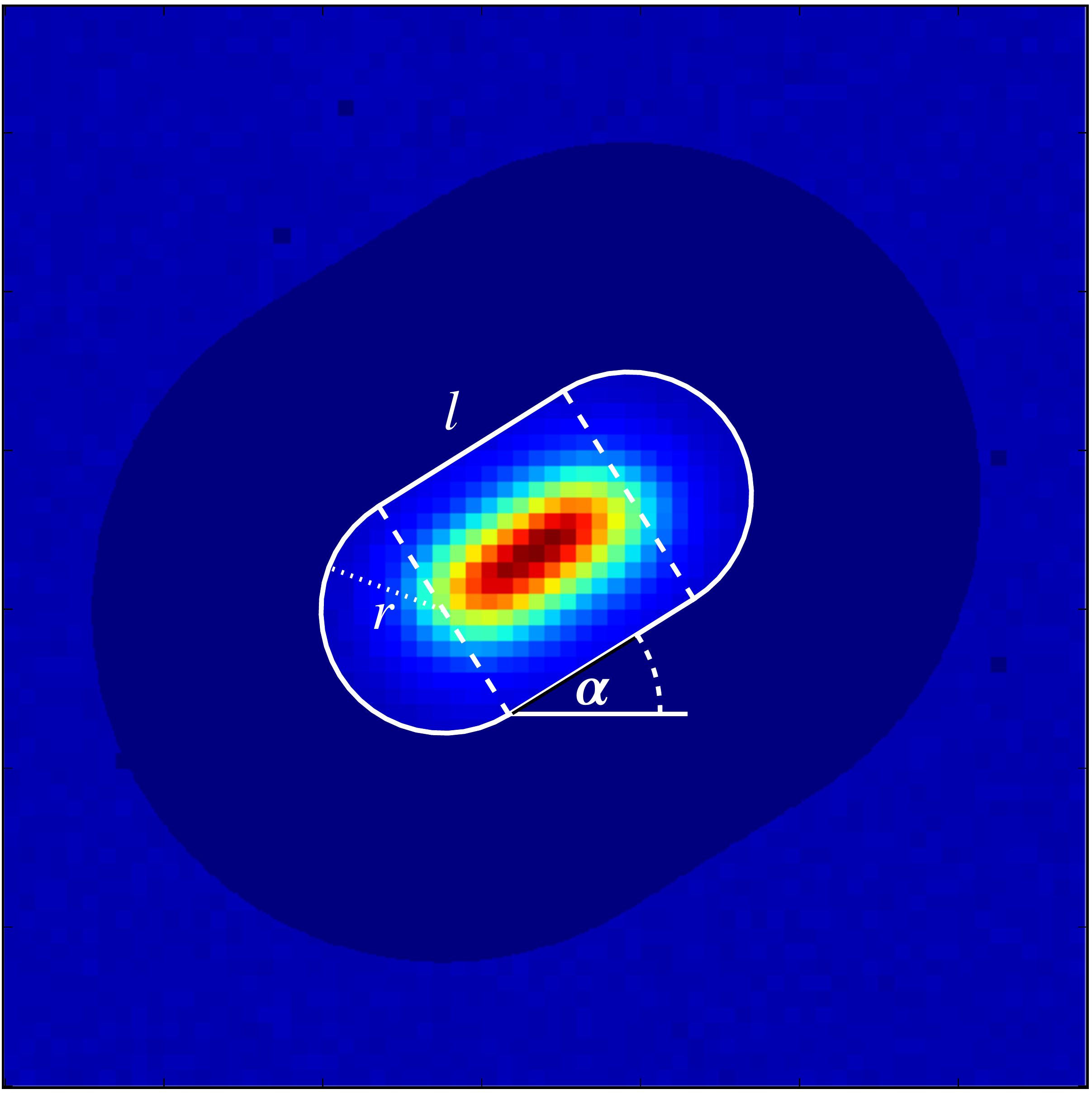}
\caption{\label{fig:pill} Example of a pill aperture. The full aperture, outlined in solid white, is the combination of a rectangle of length $l$, and width $2r$, with two semi-circular end-caps of radius $r$, all rotated at angle $\alpha$. For aperture photometry, the background is measured inside a user specified box, and outside a pill aperture with larger $r$ (but the same $l$). The image is a 480~s exposure of asteroid 2006, Polonskaya, taken on 2008-Jan-15 14:40:25 UTC when the asteroid had a rate of motion of 19.04~"/hour at angle 30.9$^\circ$. The peak pixels are nearly 9000 ADU brighter than the background.%
}
\end{center}
\end{figure}

To that end, we introduce the pill aperture. An example of this aperture is presented in Figure~\ref{fig:pill}. The pill can be described as a rectangle with two semi-circular end-caps, and is characterized by three parameters which are depicted in Figure~\ref{fig:pill}. These parameters are the trail length or how far the target has moved during an exposure, $l$, the ``radius'', $r$, which describes the half-width of the rectangle and the radius of each end-cap, and the angle of trailing on the image, $\alpha$. 

Compared with the use of a large circular aperture, the pill aperture avoids both the decrease in SNR and flux underestimation. For comparison with Figure~\ref{fig:trailing}, in the background limited case, a pill aperture with $r=2.5$~FWHM gathers $\sim99\%$ of the source flux, and compared to a large circular aperture, results in 20\% and 100\% increase in SNR for trail lengths of $l=1$~FWHM and $l=4$~FWHM, respectively. These SNR improvements jump to 100\% and 320\% for a small pill with $r=1.5$~FWHM. 

\begin{figure}[h!]
\begin{center}
\includegraphics[width=0.7\columnwidth]{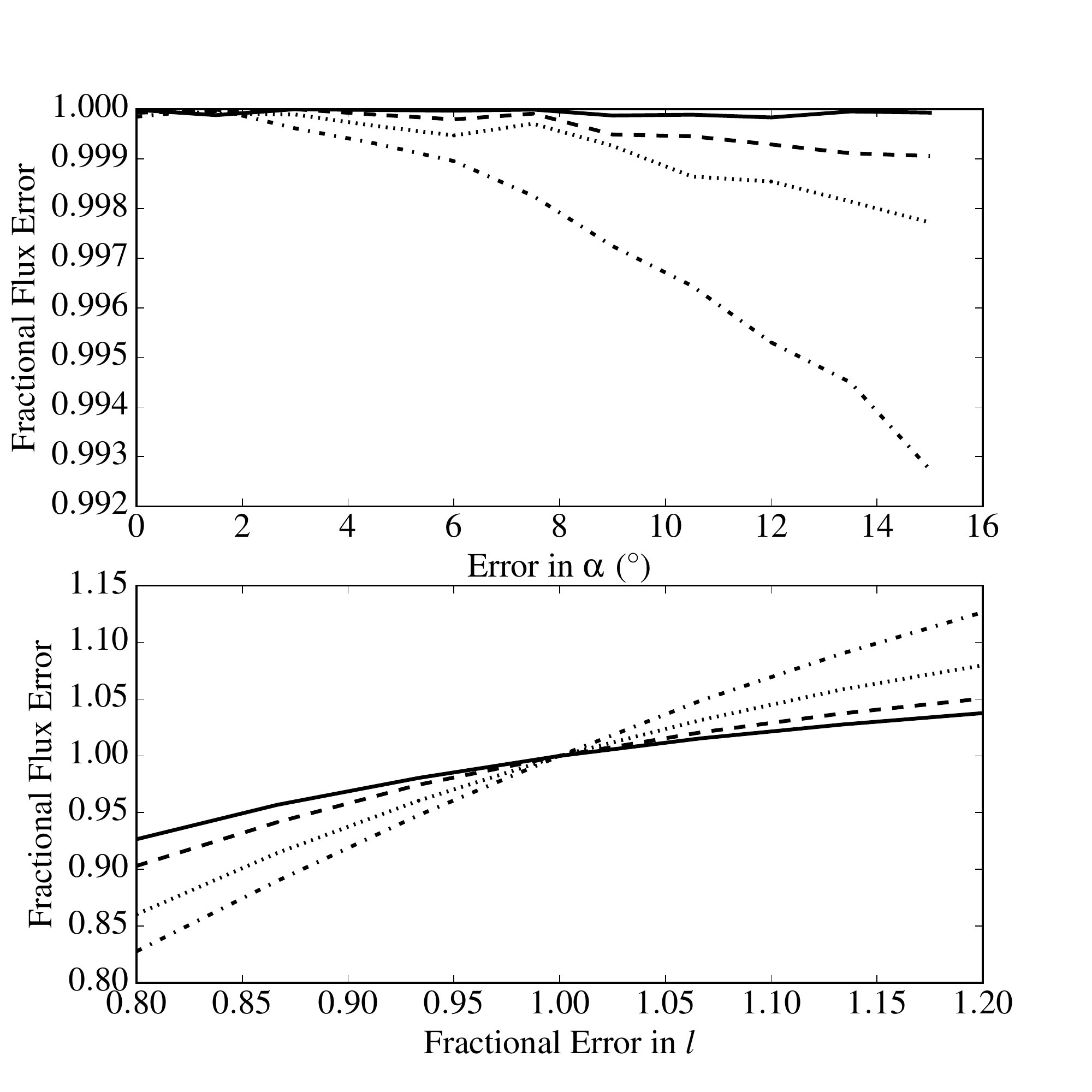}
\caption{\label{fig:fluxerror} The fractional flux error as a function of error in $l$ and $\alpha$. Curves are shown for $l$ values of 1, 2, 3, and 4 FWHM (solid, dashed, dotted, dash-dotted lines respectively). Small errors in $l$ result in significantly larger photometry errors compared to small errors in $\alpha$.%
}
\end{center}
\end{figure}

Accurate photometry clearly depends on knowledge of the trail length, $l$, and angle $r$, in an image. The effects of using incorrect values are presented in Figure~\ref{fig:fluxerror} where we plot the fraction of flux contained within an aperture with an incorrect $l$ or $\alpha$, compared to what it would be with the correct choice in $l$ and $\alpha$. These values were calculated using an aperture with $r=1.5$~FWHM and the same artificial moffat profile as used in Figure~\ref{fig:trailing}. Accurate knowledge of $l$ is more important than $\alpha$; while a 10$^\circ$ error in $\alpha$ results in a 1\% error in the measured flux, a 20\% error in $l$ results in a 20\% error in the measured flux. \citet{Veres2012} present a trail fitting technique which can be used to estimate values of $l$ and $\alpha$ when they are unknown.

\section{PSF Generation}
The pill aperture is a natural extension of the circular aperture. Once the PSF and the TSF of a source with given trail length are calculated, effective aperture corrections, or the fraction of flux inside the pill aperture as a function of radius, $r$, can be calculated directly from the TSF. Sufficiently accurate knowledge of the TSF can be found from convolution of the PSF to allow precise aperture corrections to be estimated. Even for highly trailed asteroids, aperture corrections with error $\lesssim0.5\%$ can be calculated using the PSF and TSF generation procedure we describe here.

Starting from a list of well isolated, unsaturated point sources, the PSF and TSF are found by:
\begin{enumerate}
\item super-sampling each source by some sampling factor, $5\leq ssf\leq 10$
\item fitting a moffat profile to each super sampled source and determining the mean profile for the full image
\item subtracting the mean moffat profile from the super-sampled image
\item using piece-wise constant interpolation, shifting each super-sampled residual image to account for sub-pixel centroid differences
\item generating a 2 dimensional mean lookup table as the mean of all shifted, moffat-subtracted images
\item creating a normalized, super-sampled image of the mean moffat profile, and adding the lookup table
\item convolving both the super-sampled moffat image and mean lookup table with an image of a line with trail length, $l$, and angle, $\alpha$, equal to that of the trailed source in question. 
\end{enumerate}

The PSF then consists of the mean moffat profile (found in step 2) summed with the lookup table (generated in step 5), down-sampled to the pixel scale of the original image. The estimated TSF for a source with given $\alpha$ and $l$ is then just the down-sampled, convolved image generated in step 7. Step 7 can then be repeated for different combinations of $l$ and $\alpha$ to generate TSFs for each moving source. Finally, aperture corrections can be calculated directly from the TSF.

This procedure is reminiscent of the reference standard {\it daophot - iraf} procedure with notable differences being that {\it daophot} utilizes linear interpolation to account for sub-pixel centroid differences (our step 4), and utilizes a SNR weighting scheme and $ssf=2$ during lookup table generation (our step 5) \citep{Stetson1990}. We have experimented with all of these changes, and note only minor differences in PSF and TSF quality. As noted below however, our standard procedure, with $ssf=10$ slightly out performs the {\it daophot} package when the ``worst residual'' pixel is used as the performance metric. Alternatively, in a chi-squared sense, {\it daophot} slightly out performs our routine achieving a few percent smaller chi-squared values. 

\subsection{Test of the Procedure}
\begin{figure}[h!]
\begin{center}
\includegraphics[width=0.7\columnwidth]{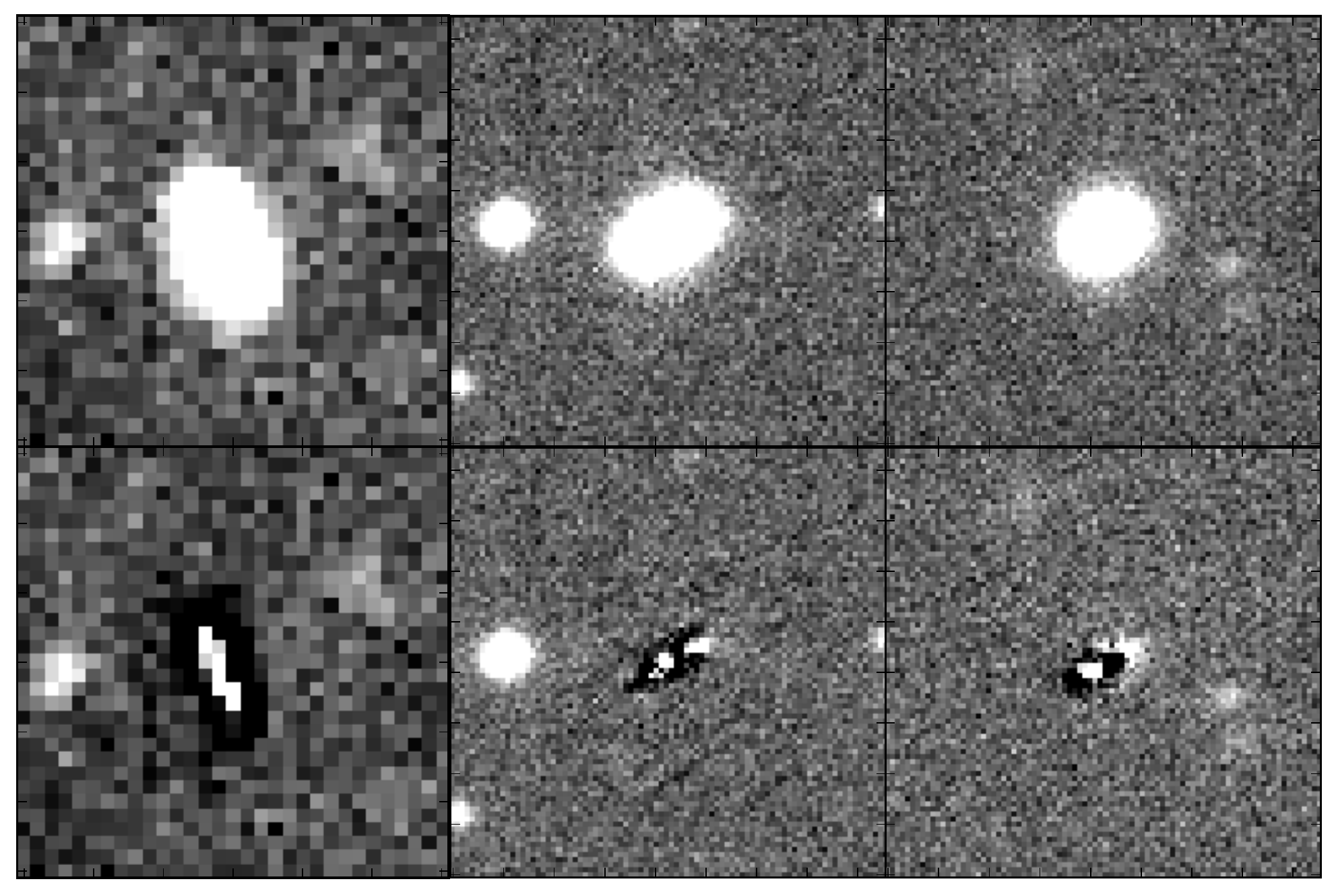}
\caption{\label{fig:TSF} Example image cutouts of asteroid 105217 (left) and Polonskaya (480~s, r' - left, 240~s g' - right) in the top row, and the residual after removal of each image's TSF in the bottom row. The asteroid is the bright target in the center of the top three images. Images are sorted left to right in order of decreasing trail length. Trail lengths are 3.91" (3.0 FWHM), 2.54" (2.3 FWHM), and 1.27" (1.3 FWHM). TSF subtractions result in residual pixels of $<10\%$, $<8\%$, and $<5\%$ the peak pixels of each asteroid's image. All residual images contain subtle structure not represented in the stationary source PSF.
}
\end{center}
\end{figure}

We test the precision of this process using 2 archival images from the Mosaic2 camera on the Blanco telescope of asteroid 105217 \citep{Elliot2005}, and 4 archival images from MegaCam on the Canada-France-Hawaii Telescope that serendipitously imaged the bright asteroid (2006) Polonskaya \citep{Gwyn2012}. Two images of Polonskaya were available in each of the r' and g' filters, with exposure times of 480 and 240~s, respectively. During the observations, the seeing was $\sim1.1$" (nearly 6 pixels), and Polonskaya was moving at 30-31"/hr. During the 240~s images of 105217 which were acquired in the VR filter, the asteroid had rate of motion $\sim~58.7$~"/hr and the seeing was $\sim1.3"$ (2.6 pixels). The images of Polonskaya resulted in a SNR$\sim800$ while those of 105217 resulted in a $SNR\sim350$  when measured in a pill aperture with $r=1.4\mbox{ FWHM}$. The long trails and extremely high SNR presents both a strong challenge, and excellent test case for our TSF generation procedure. Three of six images are shown in Figure~\ref{fig:TSF}, and a pill aperture using Polonskaya's rate of motion can be found in Figure~\ref{fig:pill}. 

For all 6 images, the PSF and the TSF were generated using the above procedure with a super-sampling factor, $ssf=10$. On-sky rates of motion for each image were provided by the JPL horizons ephemeris service\footnote{\url{http://ssd.jpl.nasa.gov/?horizons}}. Point sources were visually confirmed to be isolated stars, and only those with $SNR>200$ (measured in a circular aperture) were used. We note however, that very little difference in performance was found when considering reference star SNR thresholds in the 100-300 range. That is, the worst residual value did not change by more than a few percent.

Source centroids and SNR estimates were measured with SExtractor \citep{Bertin1996}. We tested possible improvement with measuring centroids during the moffat profile fitting (step 2). No noticeable improvement was found, and the centroids provided by SExtractor were deemed as sufficient for our purposes.

\begin{figure}[h!]
\begin{center}
\includegraphics[width=0.7\columnwidth]{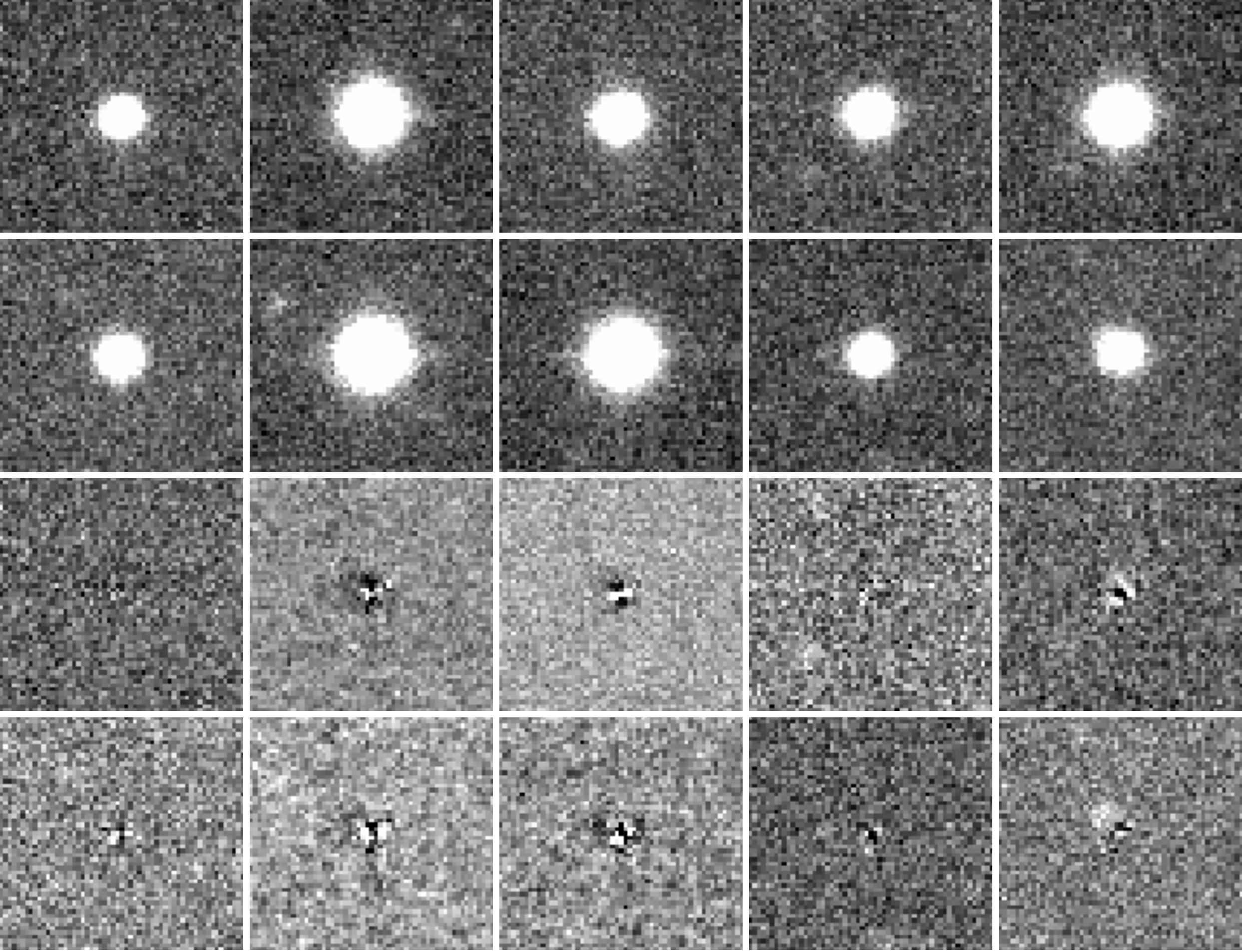}
\caption{\label{fig:PSF} The stars used for PSF generation of the r' image of Polonskaya shown in Figure~\ref{fig:TSF} (top 2 rows), and their corresponding residuals after PSF removal (bottom two rows). This is the same image as presented in the top-right panel of Figure~\ref{fig:TSF}. For each star, pixel residuals in the core of each star are $<1\%$ the peak brightness of that star. All images are brightness scaled using z-scale.%
}
\end{center}
\end{figure}

The quality of the PSF and TSF generated with $ssf=10$ are presented in Figures~ \ref{fig:TSF} and \ref{fig:PSF}. The stellar residuals of those stars used to generate the PSF are particularly good, with no consistent residual structure between different sources, and peak residuals $<1\%$ of the peak pixel amplitudes of each source.\footnote{We choose to discuss the quality of our model with pixel residuals rather than the usual chi-squared approach, as the latter implicitly assumes one is using the correct model to describe the sources. Since we make use of a lookup table, this assumption is violated, and use of a chi-squared metric is misguided.} This performance slightly surpasses that of the {\it daophot-iraf} package, which produces peak residuals of $\sim2\%$ using identical PSF reference stars, albeit with slightly worse chi-squared. Our residuals increase by a factor of $\sim2$ if 5x super sampling is used, and by $\sim5$ if 3x is used.

For the asteroids, the residuals do not fare as well; the pixel residuals are $<5-10\%$ the peak. Further, each image shows structure that shares an axis with the trailing direction. These structures dominate the residuals at all super-sampling factors between 3 and 30, and are simply structure in the Polonskaya image that are not accounted for by our TSF. 

The most likely cause of the residual structure is the inherent assumption by our routine of perfectly stable conditions during an exposure. That is, for our method to work perfectly, there can be zero variation in the PSF during the entire exposure, a condition that is impossible to meet under even the best observing circumstances. We provide basic tests of this scenario, by planting artificial trailed sources in our images containing Polonskaya, and then subtracting a source with the same flux, but slightly adjusted FWHM. These experiments revealed that $\sim 8\%$ of peak residuals are found with FWHM variations as little as $5\%$. With this result in mind, residual structures in trailed image subtractions are an inevitable consequence of the fact that point sources do not preserve a record of on-image seeing variations during an exposure. It should be noted however, that trailed subtraction residuals will be lower for less trailed sources; the 240~s exposures of Polonskaya have a peak residual nearly half the 8\% residual of the 480~s exposures, and the longer trailed image 105217 has higher peak residual still at 10\% of peak pixel value.

\begin{figure}[h!]
\begin{center}
\includegraphics[width=0.7\columnwidth]{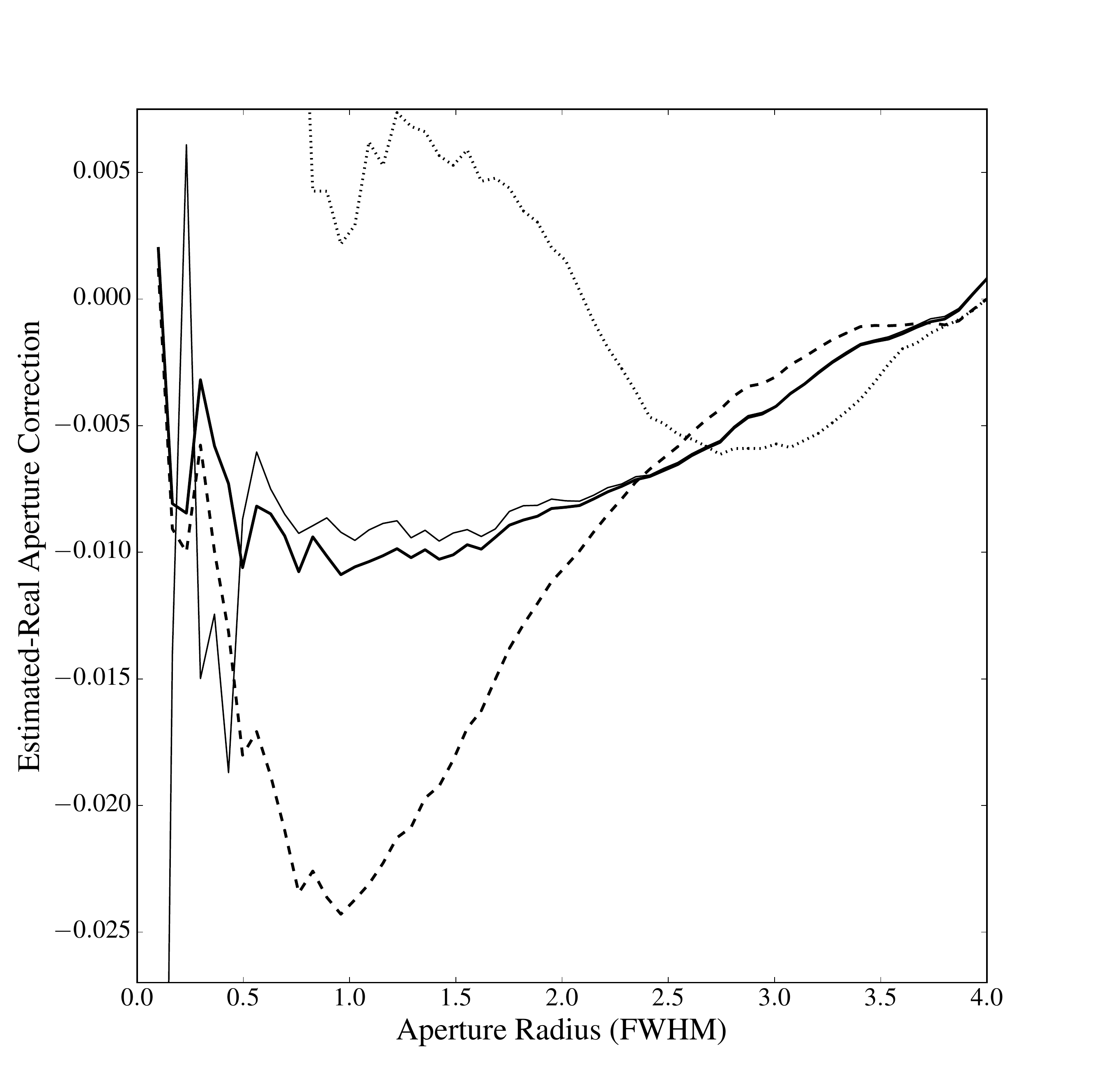}
\caption{\label{fig:AperCorrDiff} Difference in aperture correction measured from the asteroid images directly vs. that estimated from the TSF of 105217 (ssf=10, dotted line), of Polonskaya (ssf=5 solid thin line, and ssf=10 solid thick line) and of Polonskaya without the use of the lookup table (dashed line). If the lookup table is utilized, accurate aperture corrections can be calculated directly from the TSF, even for highly trailed sources.%
}
\end{center}
\end{figure}

Despite the modest subtraction residuals, accurate photometry for all trailed sources is still afforded by the TSF. This is because accurate aperture corrections can be measured directly from the TSF. We demonstrate this with the extremely high SNR images of Polonskaya and 105217. Specifically, we measured the aperture corrections as a function of $r$ in the standard way, using the TSF, and then directly from the image of the asteroids themselves. The difference is presented in Figure~\ref{fig:AperCorrDiff} for one of the 480~s, r' images of Polonskaya, and for one image of 105217. For completeness, we also include an estimate of the aperture correction when the lookup table is not utilized. As can be seen, for nearly all $r$, the difference between the aperture corrections measured from the TSF, and that measured directly from the asteroid's image is less than 10 millimags. In a $r=1.4$~FWHM pill aperture, the $SNR=889$ corresponds to a source photon noise uncertainty of $\sim1$~millimag. Clearly, even for highly elongated sources, accurate aperture corrections can be measured from the TSF alone. It should be noted that with a pill radii $r<2.5$, for the extremely high SNR images of Polonskaya, the aperture correction uncertainty is the dominant source of error, and a pill aperture with larger $r$ should be used. For images with SNR$<200$ (or still in the flux limited regime) however, the uncertainty is dominated by source photon noise, and apertures of smaller $r$ will provide the lowest final uncertainty. The value of $r$ that provides maximal SNR depends on the trail length, but we have found in practice that for background images with trail length $<2$~FWHM, $1.2<r<1.6$ typically provide maximal SNR.

In Figure~\ref{fig:AperCorrDiff}, we also present the results when a lookup table is not used. This facilitates comparison with the trail-fitting method of \citet{Veres2012} which utilizes a Gaussian profile without a lookup table, but otherwise, uses essentially the same technique to generate TSFs as we present here. Without the lookup table, the error in the aperture correction estimate is a factor of 4 worse. Therefore, For the most accurate photometry, the use of a lookup table is required. For sources with SNR$\geq50$, the lack of a lookup table will cause a bias in the flux measurement larger than the amplitude of the photon noise uncertainty.

Recently, \citet{Sonnett2013} presented a comparison of the relative performances of various photometry techniques in the scenario of moderate SNR observations ($\gtrsim20$), of a Kuiper Belt Object, moving at 2.8 "/hr. At the expense of knowledge of the PSF and any aperture corrections, the telescope was tracked at a non-sidereal rate to produce a (nearly) untrailed image of the KBO, allowing circular apertures to be used, facilitating comparison of a wide range of photometry packages.

In a lowest RMS sense, one of the highest performing techniques is the \emph{SExtractor} \emph{MAG\_ISOCOR} photometry technique. This technique attempts to provide full flux measurements of a source by tracing an isophotal curve around the source and measuring photometry within that aperture. A correction is then applied to account for the flux excluded by that aperture based on a 2 dimensional Gaussian profile fit to the source. In this way, source elongation can be at least roughly accounted for, with the RMS uncertainty increasing by modest amounts - $\sim50\%$ in the case discussed by \citet{Sonnett2013}.

\begin{figure}[h!]
\begin{center}
\includegraphics[width=0.7\columnwidth]{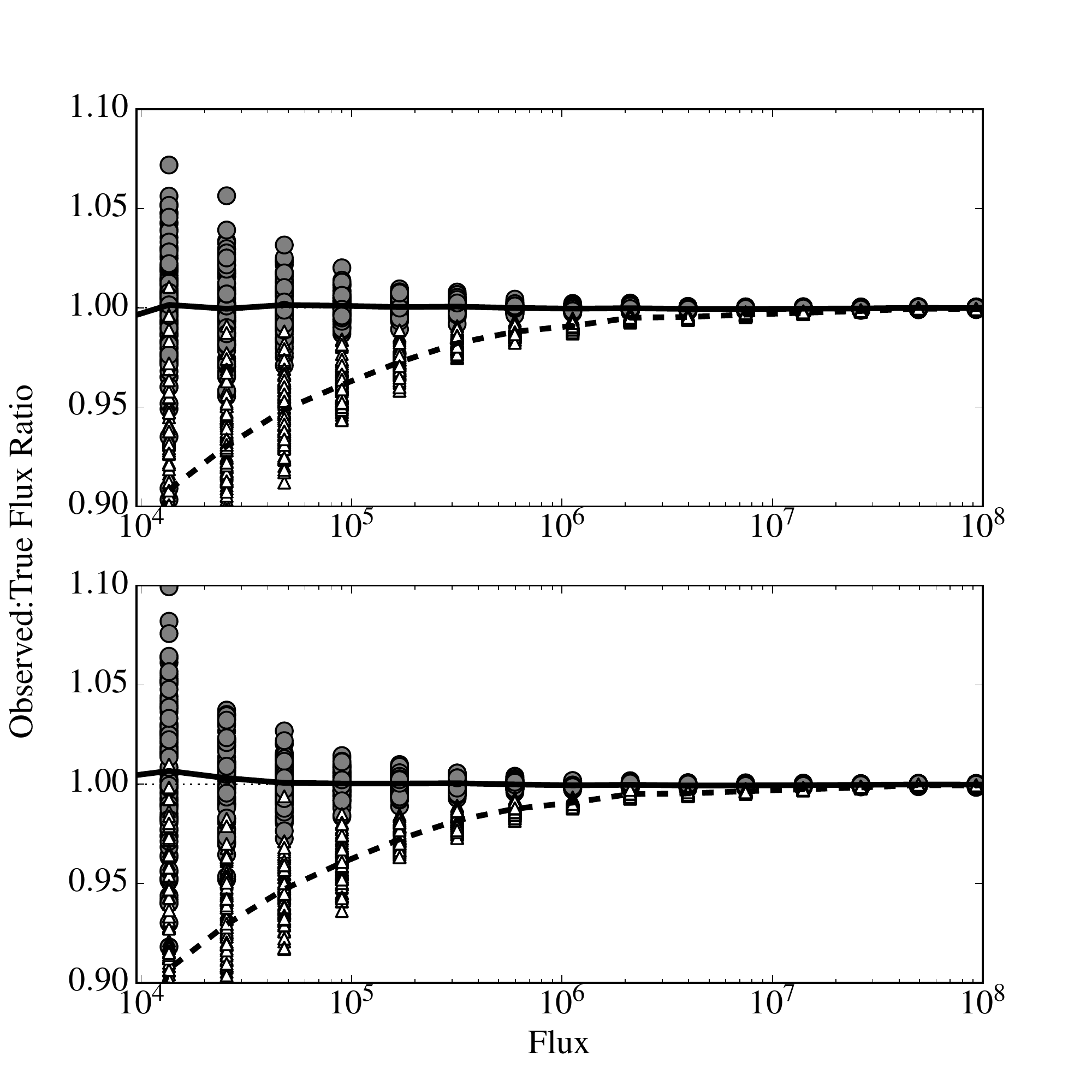}
\caption{\label{fig:magisocor} Observed:true flux ratio of simulated trailed sources with trail lengths 0.9 (top) and 2.5 FWHM (bottom). Grey circles are aperture-corrected pill photometry of the source, white triangles are the photometry reported by \emph{SExtractor} \emph{MAG\_ISOCOR}. The mean photometric values at each flux level are displayed by the solid and dashed lines for the pill aperture and \emph{MAG\_ISOCOR} techniques, respectively. Both techniques produce similar RMS scatter at each flux level (for reference, $0.015$ magnitudes at $10^5$ flux level). \emph{MAG\_ISOCOR} however, clearly results in a mean flux level which is biased to low values at low SNR.%
}
\end{center}
\end{figure}

One large difficulty arises in the use of \emph{MAG\_ISOCOR} when tracing isophotal apertures around low SNR sources. While the SNR of this technique is generally as good or better than other techniques, a bias towards measuring a flux lower than the true value exists. This is presented in Figure~\ref{fig:magisocor} where we compare simulations of photometry reported by pill aperture photometry and that from \emph{MAG\_ISOCOR}. This comparison was made by planting 50 synthetically elongated Moffat profiles at various flux levels in an image with a background of 1000 ADU, and appropriate Gaussian noise. The Moffat profile extracted from the first r' image of asteroid Polonskaya was used. The rate, and 1/3rd the rate of motion of the asteroid in that image were considered. The pill aperture photometry was corrected for the aperture correction measured on frame to provide comparison with \emph{MAG\_ISOCOR}, which includes their own corrections. All measurements at each flux level are normalized by the true flux, and are presented in Figure~\ref{fig:magisocor}, along with the mean measurement at each flux level.

As can be seen, both produce similar SNRs at all flux levels. Similiarly, at high SNR levels, both techniques reproduce the correct flux with no appreciable biases. At low SNR however, on average, \emph{MAG\_ISOCOR} produces a mean measurement that is biased faintward while the pill aperture produces an unbiased measurement. For example, at SNR~$\sim100$ the \emph{MAG\_ISOCOR} bias rapidly degrades to $\sim3\%$ at SNR~$\sim50$, a bias which is larger than the uncertainty at that SNR. The reason pill photometry avoids this bias is simple; measurement of the PSF, and estimation of the TSF afforded by sidereally tracked images allows accurate aperture corrections to be calculated, at least to the precision demonstrated in Figure~\ref{fig:AperCorrDiff}.

\subsection{Background Estimation}
In addition to the pill aperture, we also introduce a new technique to estimate the background modal value in an image. A common technique used to estimate the mode is to take 3 three times the median value minus 2 times the mean value of a sourceless region of an image, or small variations on this equation \citep[see for example][]{Bertin1996}. Our modal  technique involves multiplying the pixel values within a region by some factor $x<1$ and rounding it to an integer to produce values $y_\textrm{i}$. The mode $m$ of all $y_\textrm{i}$ integers is then found. The background estimate is then the median value of all pixels with $int\left(x\times y_{\textrm{i}}\right)=m$. Our tests demonstrate that for values of $x\sim0.1$, the background estimated by our modal technique is less sensitive to bright background sources than the more standard background estimation techniques. 

\begin{figure}[h!]
\begin{center}
\includegraphics[width=0.7\columnwidth]{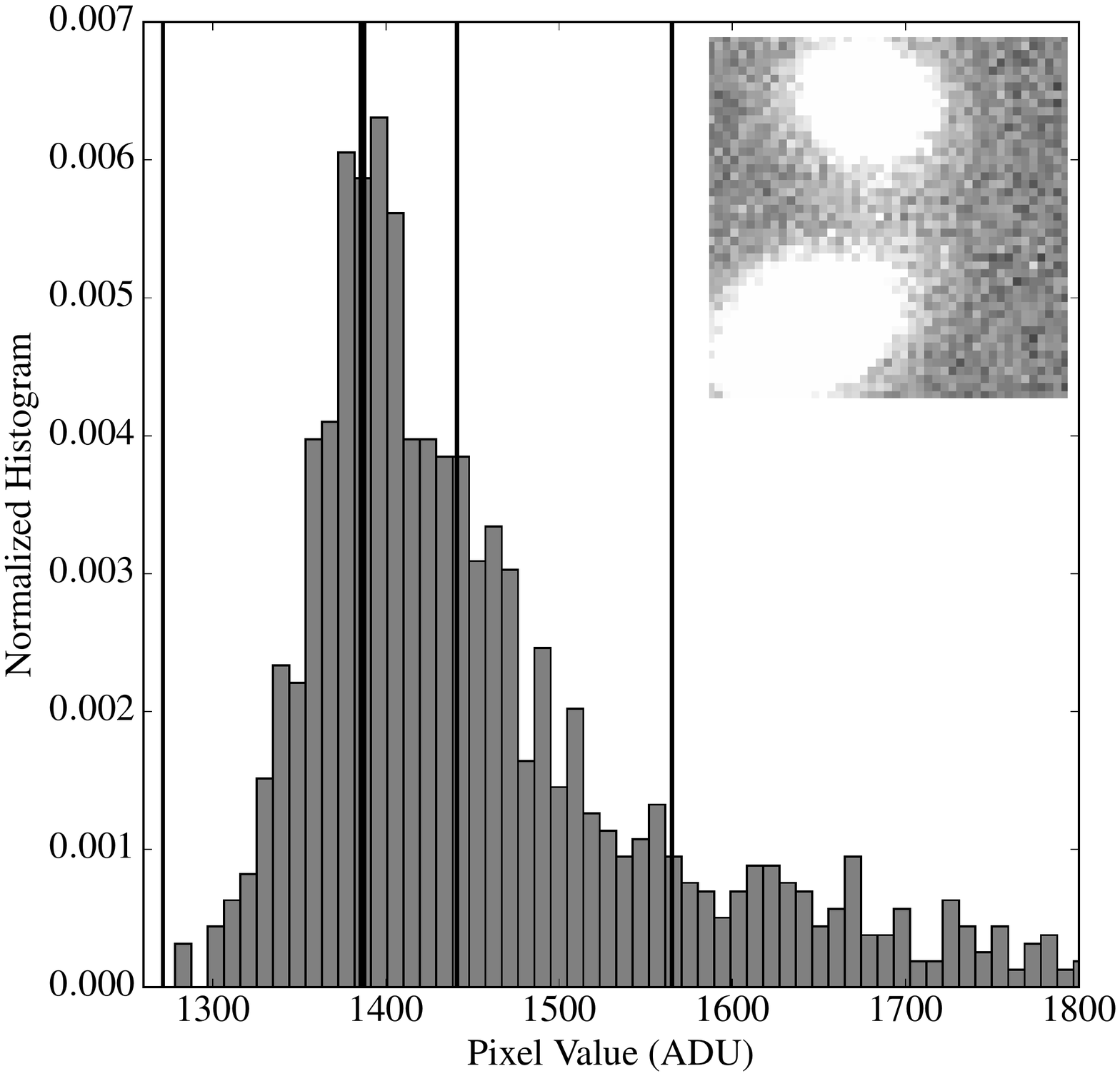}
\caption{\label{fig:bg} A mormalized pixel histogram of the 41x41 pixel image displayed in the inset. Vertical lines represent background estimates from different techniques. From left to right those techniques are: $3\times\textrm{median}-2\times\textrm{mean}$ with an initial 3-sigma standard deviation rejection away from the median; modal estimate with x=0.1 and x=0.2; median; and mean. 15\% of the image (260 pixels) corresponding to the centres of the foreground galaxy and star have pixel values as high as 3260 ADU and are beyond the right limit of the plot. The modal technique clearly reports the most reasonable background estimate in face of bright foreground source contamination.%
}
\end{center}
\end{figure}

A comparison of techniques is demonstrated in Figure~\ref{fig:bg} where we have cut out a small 41x41 pixel region specifically chosen to contain two bright sources which occupy roughly half of the cutout. We present a histogram of pixel values, and background estimations using various techniques, including mean, median, Gaussian fit (which produces the same value as the mean), $3\times \textrm{median}-2\times\textrm{mean}$ with initial 3-sigma away from the mean rejection, and our modal technique using $x=0.1$ and $x=0.2$. As can be seen, the modal technique produces visually the most satisfactory result. 

It should be noted that the modal technique does depend on the choice in $x$. Our experience suggests that $x\gtrsim0.3$, while still outperforming other techniques, produces background estimates that are a few percent too low, while values $x\lesssim0.05$ produce values that are too large unless the region is large, containing of order 10,000 or more pixels. We consider $x=0.1$ to be the best overall choice.

\section{Conclusions}
We introduce the pill aperture, a natural extension of the circular aperture for photometry of trailed sources. Described by a trail length, radius, and trail angle, the pill aperture allows increased precision photometry by minimizing the number of background pixels included in the aperture, and allows the majority of source flux to be included in the aperture without the use of unnecessarily large circular apertures. 

In this paper we presented a new method to generate both the point spread function (PSF) and the point spread function of trailed sources (TSF) from the PSF. From the TSF, accurate aperture corrections can be estimated for a given trailed source, allowing for the use of small apertures, which maximize SNR within the aperture. We found that for pill apertures with radii at least as large as the FWHM, the error in aperture correction is less than 10 millimags for a long 480~s exposure of fast moving asteroid, Polonskaya. The software for pill aperture photometry, and PSF+TSF generation are available in the python package TRIPPy.

\section{Acknowledgements}

This research made use of Astropy, a community-developed core Python package for Astronomy (Astropy Collaboration, 2013). This research also used the facilities of the Canadian Astronomy Data Centre operated by the National Research Council of Canada with the support of the Canadian Space Agency.

This research made use of the Giorgini, JD and JPL Solar System Dynamics Group, NASA/JPL Horizons On-Line Ephemeris System, \url{http://ssd.jpl.nasa.gov/?horizons}, from which data were retrieved in December, 2015.

This research made use of PyFits, a product of the Space Telescope Science Institute, which is operated by AURA for NASA.

Software: \software{IRAF}, \software{SExtractor}, \software{SciPy \citep{Oliphant_2007}}, \software{Matplotlib Hunt\citep{Hunter_2007}}, \software{PyFits}

\nocite{*}

\bibliographystyle{aasjournal}
\bibliography{converted_to_latex}

\end{document}